\documentclass[twoside,twocolumn,9pt]{article}
\usepackage{extsizes}
\usepackage[super,sort&compress,comma]{natbib} 
\usepackage[version=3]{mhchem}
\usepackage[left=1.5cm, right=1.5cm, top=1.785cm, bottom=2.0cm]{geometry}
\usepackage{balance}
\usepackage{mathptmx}
\usepackage{sectsty}
\usepackage{graphicx} 
\usepackage{lastpage}
\usepackage[format=plain,justification=justified,singlelinecheck=false,font={stretch=1.125,small,sf},labelfont=bf,labelsep=space]{caption}
\usepackage{float}
\usepackage{fancyhdr}
\usepackage{fnpos}
\usepackage[english]{babel}
\addto{\captionsenglish}{%
  
}
\usepackage{array}
\usepackage{droidsans}
\usepackage{charter}
\usepackage[T1]{fontenc}
\usepackage[usenames,dvipsnames]{xcolor}
\usepackage{setspace}
\usepackage[compact]{titlesec}
\usepackage{hyperref}
\usepackage[export]{adjustbox}

\usepackage{epstopdf}%This line makes .eps figures into .pdf - please comment out if not required.

\definecolor{cream}{RGB}{222,217,201}

\begin{document}

\pagestyle{fancy}
\thispagestyle{plain}
\fancypagestyle{plain}{
%%%HEADER%%%
\renewcommand{\headrulewidth}{0pt}
}
%%%END OF HEADER%%%

\newcommand\fs{\mbox{$.\!\!^{\mathrm s}$}}% 
\newcommand\sq{\mbox{\rlap{$\sqcap$}$\sqcup$}}% 
\newcommand\arcdeg{\mbox{$^\circ$}}% 
\newcommand\arcmin{\mbox{$^\prime$}}% 
\newcommand\arcsec{\mbox{$^{\prime\prime}$}}% 
\newcommand\farcs{\mbox{$.\!\!^{\prime\prime}$}}
\newcommand\degr{\arcdeg}% 

\newcommand\aj{{AJ}}%        % Astronomical Journal 
\newcommand\psj{{PSJ}}%       % Planetary Science Journal
\newcommand\araa{{ARA\&A}}%  % Annual Review of Astron and Astrophys 
\newcommand\apj{{ApJ}}%    % Astrophysical Journal 
\newcommand\apjl{{ApJL}}     % Astrophysical Journal, Letters 
\newcommand\apjs{{ApJS}}%    % Astrophysical Journal, Supplement 
\newcommand\aap{{A\&A}}%     % Astronomy and Astrophysics 
\newcommand\aapr{{A\&A~Rv}}%  % Astronomy and Astrophysics Reviews 
\newcommand\aaps{{A\&AS}}%    % Astronomy and Astrophysics, 
\newcommand\mnras{{MNRAS}}%   % Monthly Notices of the RAS 
\newcommand\physrep{{Phys.~Rep.}}
\newcommand\nat{{Nature}}
\newcommand\jqsrt{J.~Quant.~Spec.~Radiat.~Transf.} 
\newcommand\jcp{J. Chem. Phys.} 
\let\astap=\aap 
\let\apjlett=\apjl 
\let\apjsupp=\apjs 
\let\applopt=\ao

%%%PAGE SETUP - Please do not change any commands within this section%%%
\makeFNbottom
\makeatletter
\renewcommand\LARGE{\@setfontsize\LARGE{15pt}{17}}
\renewcommand\Large{\@setfontsize\Large{12pt}{14}}
\renewcommand\large{\@setfontsize\large{10pt}{12}}
\renewcommand\footnotesize{\@setfontsize\footnotesize{7pt}{10}}
\renewcommand\scriptsize{\@setfontsize\scriptsize{7pt}{7}}
\makeatother

\renewcommand{\thefootnote}{\fnsymbol{footnote}}
\renewcommand\footnoterule{\vspace*{1pt}% 
\color{cream}\hrule width 3.5in height 0.4pt \color{black} \vspace*{5pt}} 
\setcounter{secnumdepth}{5}

\makeatletter 
\renewcommand\@biblabel[1]{#1}            
\renewcommand\@makefntext[1]% 
{\noindent\makebox[0pt][r]{\@thefnmark\,}#1}
\makeatother 
\renewcommand{\figurename}{\small{Fig.}~}
\sectionfont{\sffamily\Large}
\subsectionfont{\normalsize}
\subsubsectionfont{\bf}
\setstretch{1.125} %In particular, please do not alter this line.
\setlength{\skip\footins}{0.8cm}
\setlength{\footnotesep}{0.25cm}
\setlength{\jot}{10pt}
\titlespacing*{\section}{0pt}{4pt}{4pt}
\titlespacing*{\subsection}{0pt}{15pt}{1pt}
%%%END OF PAGE SETUP%%%

%%%FOOTER%%%
\fancyfoot{}
\fancyfoot[LO,RE]{\vspace{-7.1pt}\includegraphics[height=9pt]{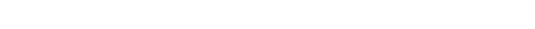}}
\fancyfoot[CO]{\vspace{-7.1pt}\hspace{13.2cm}\includegraphics{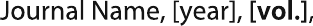}}
\fancyfoot[CE]{\vspace{-7.2pt}\hspace{-14.2cm}\includegraphics{head_foot/RF}}
\fancyfoot[RO]{\footnotesize{\sffamily{1--\pageref{LastPage} ~\textbar  \hspace{2pt}\thepage}}}
\fancyfoot[LE]{\footnotesize{\sffamily{\thepage~\textbar\hspace{3.45cm} 1--\pageref{LastPage}}}}
\fancyhead{}
\renewcommand{\headrulewidth}{0pt} 
\renewcommand{\footrulewidth}{0pt}
\setlength{\arrayrulewidth}{1pt}
\setlength{\columnsep}{6.5mm}
\setlength\bibsep{1pt}
%%%END OF FOOTER%%%

%%%FIGURE SETUP - please do not change any commands within this section%%%
\makeatletter 
\newlength{\figrulesep} 
\setlength{\figrulesep}{0.5\textfloatsep} 

\newcommand{\topfigrule}{\vspace*{-1pt}% 
\noindent{\color{cream}\rule[-\figrulesep]{\columnwidth}{1.5pt}} }

\newcommand{\botfigrule}{\vspace*{-2pt}% 
\noindent{\color{cream}\rule[\figrulesep]{\columnwidth}{1.5pt}} }

\newcommand{\dblfigrule}{\vspace*{-1pt}% 
\noindent{\color{cream}\rule[-\figrulesep]{\textwidth}{1.5pt}} }

\makeatother
%%%END OF FIGURE SETUP%%%

%%%TITLE AND AUTHORS%%%
\twocolumn[
  \begin{@twocolumnfalse}
{\includegraphics[height=30pt]{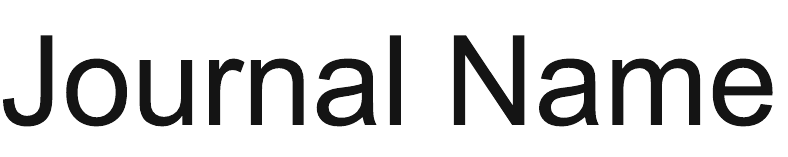}\hfill\raisebox{0pt}[0pt][0pt]{\includegraphics[height=55pt]{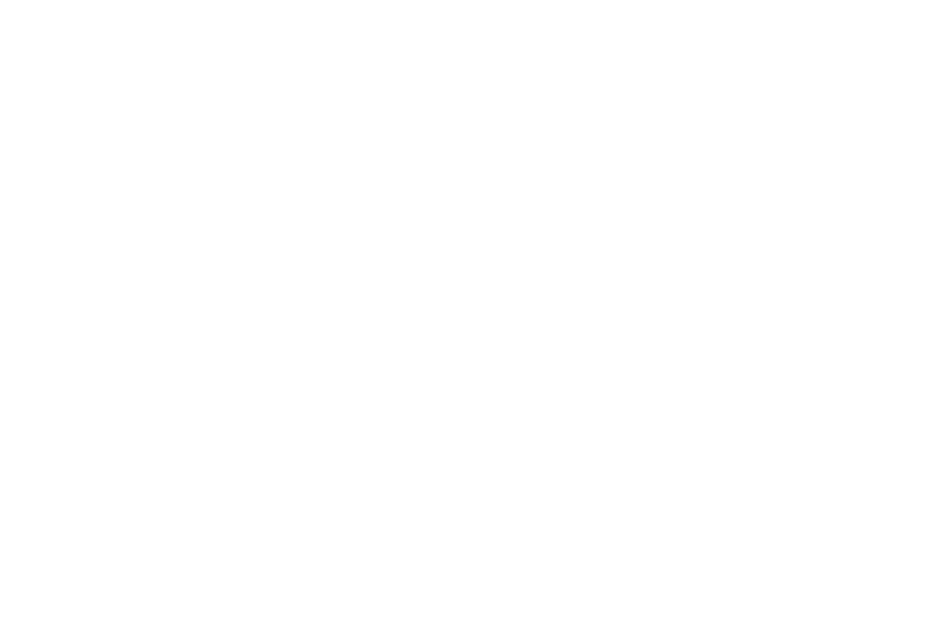}}\\[1ex]
\includegraphics[width=18.5cm]{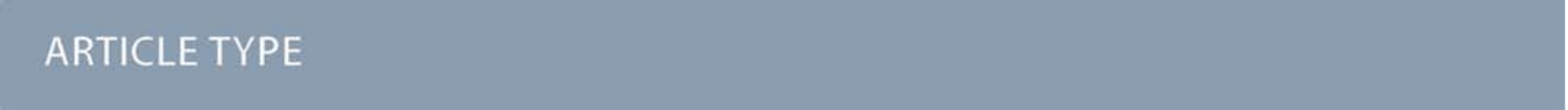}}\par
\vspace{1em}
\sffamily
\begin{tabular}{m{4.5cm} p{13.5cm} }

\includegraphics{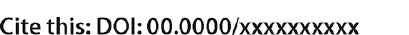} & \noindent\LARGE{Streamers feeding the SVS13-A protobinary system: astrochemistry reveals accretion shocks?}

\\%Article title goes here instead of the text "This is the title"
 & \vspace{0.3cm} \\

 & \noindent\large{Eleonora Bianchi,\textit{$^{a}$} Ana López-Sepulcre,\textit{$^{b,c}$} Cecilia Ceccarelli,\textit{$^{b}$} Claudio Codella,\textit{$^{d,b}$} Linda Podio,\textit{$^{d}$} Mathilde Bouvier,\textit{$^{e}$} Joan Enrique-Romero,\textit{$^{f}$} Rafael Bachiller,\textit{$^{g}$} and Bertrand Lefloch\textit{$^{b,h}$}}\\ 
 %and Francesco Fontani\textit{$^{d}$}}\\%Author names go here instead of "Full name", etc.

\includegraphics{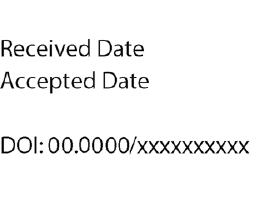} & \\

\end{tabular}

 \end{@twocolumnfalse} \vspace{0.6cm}

  ]
%%%END OF TITLE AND AUTHORS%%%

%%%FONT SETUP - please do not change any commands within this section
\renewcommand*\rmdefault{bch}\normalfont\upshape
\rmfamily
\section*{}
\vspace{-1cm}

%%%FOOTNOTES%%%

\footnotetext{\textit{$^{a}$Excellence Cluster ORIGINS, Boltzmannstraße 2, 85748, Garching bei München, Germany. E-mail: eleonora.bianchi@origins-cluster.de}}

\footnotetext{\textit{$^{b}$Univ. Grenoble Alpes, CNRS, IPAG, 38000 Grenoble, France.}}
\footnotetext{\textit{$^{c}$Institut de Radioastronomie Millimétrique, 38406 Saint-Martin d’ Hères, France.}}
\footnotetext{\textit{$^{d}$INAF, Osservatorio Astrofisico di Arcetri, Largo E. Fermi 5, I-50125, Firenze, Italy.}}
\footnotetext{\textit{$^{e}$Leiden Observatory, Leiden University, P.O. Box 9513, 2300 RA, Leiden, The Netherlands}}
\footnotetext{\textit{$^{f}$ Leiden Institute of Chemistry, Gorlaeus Laboratories, Leiden University, P.O. Box 9502, 2300 RA Leiden, The Netherlands}}
\footnotetext{\textit{$^{g}$Observatorio Astronómico Nacional (OAN-IGN), Alfonso XII 3, 28014 Madrid, Spain}}
\footnotetext{\textit{$^{h}$Laboratoire d’astrophysique de Bordeaux, Univ. Bordeaux, CNRS, B18N, allée Geoffroy Saint-Hilaire, 33615 Pessac, France}}

%Please use \dag to cite the ESI in the main text of the article.
%If you article does not have ESI please remove the the \dag symbol from the title and the footnotetext below.
%\footnotetext{\dag~Electronic Supplementary Information (ESI) available: [details of any supplementary information available should be included here]. See DOI: 00.0000/00000000.}
%additional addresses can be cited as above using the lower-case letters, c, d, e... If all authors are from the same address, no letter is required

%\footnotetext{\ddag~Additional footnotes to the title and authors can be included \textit{e.g.}\ `Present address:' or `These authors contributed equally to this work' as above using the symbols: \ddag, \textsection, and \P. Please place the appropriate symbol next to the author's name and include a \texttt{\textbackslash footnotetext} entry in the the correct place in the list.}

%%%END OF FOOTNOTES%%%

%%%ABSTRACT%%%%

%\sffamily{\textbf{The abstract should be a single paragraph which summarises the content of the article. Any references in the abstract should be written out in full \textit{e.g.} [Surname \textit{et al., Journal Title}, 2000, \textbf{35}, 3523].}}\\%The abstrast goes here instead of the text "The abstract should be..."

We report ALMA high-angular resolution ($\sim$ 50 au) observations of the binary system SVS13-A. More specifically, we analyse deuterated water (HDO) and sulfur dioxide (SO$_2$) emission. The molecular emission is associated with both the components of the binary system, VLA4A and VLA4B. The spatial distribution is compared to that of formamide (NH$_2$CHO), previously analysed in the system\cite{Bianchi2022}.
Deuterated water reveals an additional emitting component spatially coincident with the dust accretion streamer, at a distance $\geq$ 120 au from the protostars, and at blue-shifted velocities ($>$ 3 km s$^{-1}$ from the systemic velocities). 
We investigate the origin of the molecular emission in the streamer, in light of thermal sublimation temperatures calculated using updated binding energies (BE) distributions. We propose that the observed emission is produced by an accretion shock at the interface between the accretion streamer and the disk of VLA4A. Thermal desorption is not completely excluded in case the source is actively experiencing an accretion burst.

%%%END OF ABSTRACT%%%%

\rmfamily %Please do not remove this line.

%%%MAIN TEXT%%%%
\section{Introduction}

Our understanding of how a star and a planetary system form has evolved substantially in the latest years. On the one hand, observations have revealed the presence of substructures such as gaps and rings in very young protostellar disks (age $\sim$ 10$^{5}$ years) \cite{ALMA2015,Sheehan2020,Segura-Cox2020}. On the other hand, evolved Class II disks (age $>$ 10$^{6}$ years) contain not enough dust mass to explain the observed exoplanets population, in contrast to early Class 0 and I disks which are more massive\cite{Tyc2018}.
These indications suggest that planetesimal formation start already in the early phases of protostellar disks. In this respect, it is of paramount importance to investigate the properties of young Class 0 (age $\sim$ 10$^{4}$ years) and Class I disks (age $\sim$ 10$^{5}$ years) in order to define the initial conditions for planet formation. 
Since young disks are typically deeply embedded in the parent envelope, the characterisation of their physical and chemical properties is not trivial from an observation point of view. If simple parameters, such as the disk dust and gas mass, are already difficult to measure (see e.g. \cite{Miotello2022,Sheehan2022}), their chemical composition is even less explored. Nevertheless, the molecular complexity present in the disk will be inherited from the forming planets, at least in the outer regions (e.g. \cite{Oberg2021,Ceccarelli2022} and references therein).
Shedding light on the chemical composition of young disks is the only way to investigate the initial chemical budget available for planets when they start to form.
Because of an important envelope contribution, the standard gas tracers (e.g. CO isotopologues) fail in the observation of the inner disk regions of young protostars. Other tracers, such as interstellar complex organic molecules (molecules with 6 or more atoms and based on carbon; e.g. \cite{Herbst2009,Ceccarelli2022}) demonstrated to be more effective to study the complex processes happening in the disk. However, their employment require at the same time sensitive observations and a better understanding of the chemical/physical processes that release the molecules we observe in the gas. In this context, a strong synergy between astronomical observations, quantum chemistry calculations and laboratory experiments is required.
In this paper, we present new observations performed using the Atacama Large Millimeter/submillimeter Array (ALMA) of the protobinary system SVS13-A. In Sec. \ref{sec:source} we present the source background and in Sec. \ref{sec:observations} we describe the ALMA observations. The main results of our analysis are presented in Sec. \ref{sec:results}. Section \ref{sec:discussion} discuss the possible origin of the detected molecular emission, using the most up-to-date binding energies distributions. The main conclusions are reported in Sec. \ref{sec:conclusions}.

\section{The protostellar binary system SVS13-A}\label{sec:source}

\begin{figure}[h]
\centering
  \includegraphics[height=8cm]{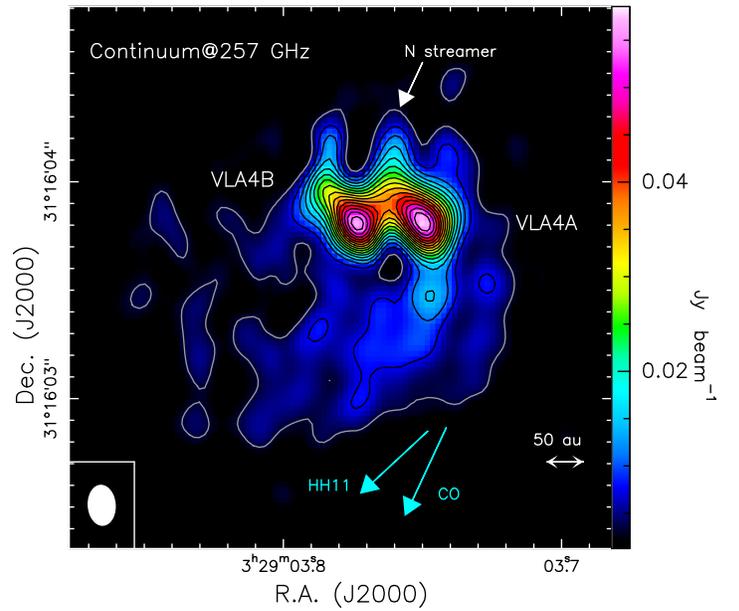}
  \caption{Continuum emission as observed by ALMA at 1.2 mm (color scale and contours) \cite{Bianchi2022}. First contours (gray) and steps (black) are 8$\sigma$, corresponding to 3.35 mJy beam$^{-1}$. The beam size is 0$\farcs$19 $\times$ 0$\farcs$13 (PA=+5$^{\circ}$). The binary system is composed of VLA4A (right) and VLA4B (left). Coordinates are $\alpha_{\rm J2000}$ = 03$^{\rm h}$ 29$^{\rm m}$ 3$\fs$75, $\delta_{\rm J2000}$ = +31$\degr$ 16$\arcmin$ 03$\farcs$81 and $\alpha_{\rm J2000}$ = 03$^{\rm h}$ 29$^{\rm m}$ 3$\fs$773, $\delta_{\rm J2000}$ = +31$\degr$ 16$\arcmin$ 03$\farcs$81 for VLA4A and VLA4B, respectively. The blue arrows indicate the directions toward HH11 (133$\degr$) as well as the main axis of the H$_{\rm 2}$/CO jet (155$\degr$) as inferred by \citet{Lefevre2017}. The white arrow indicates the position of the norther streamer discussed in the text.}
  %\textit{Right panels:} Integrated emission of one NH$_{\rm 2}$CHO and one CH$_{\rm 3}$OCH$_{\rm 3}$ line, in colour scale, superposed to the dust emission in white contours. The emission is integrated in the following velocity intervals: 4 -- 7.5 km s$^{-1}$ in the left panels (Blue), 7.5 -- 8.7 km s$^{-1}$ in the middle panels (V$_{\rm sys}$), and 8.8 -- 12.0 km s$^{-1}$ in the right panels (Red).}
  \label{fig:Continuum}
\end{figure}

SVS13-A is a perfect target to perform astrochemical studies and tackle some of the questions above.
It is very well studied, in Perseus, at a distance of $\sim$ 299 pc (\textit{Gaia}, \cite{Zucker2018}).
The system has a bolometric luminosity of 58.8 L$_{\odot}$ \cite{Fiorellino2021} and it is at the origin of the famous Herbig–Haro (HH) chain 7--11 \cite{Reipurth1993}, a molecular jet and a large-scale outflow\cite{Lefloch1998,Lefevre2017}.
SVS13-A is composed of a close binary system with a separation of 0$\farcs$3\cite{Anglada2000}. The proper motion of the two components VLA4A and VLA4B have been investigated for almost 30 yrs \cite{Carrasco2008,Diaz2022}. 
A third companion, called SVS13A2 is detected at a distance of $\sim$ 5\arcsec \cite{Rodriguez1999,Tobin2016}.
The continuum emission at high-angular resolution shows two circumstellar disks and circumbinary material distributed in spiral arms which suggest accretion streamers (see Fig. \ref{fig:Continuum} \cite{Tobin2016,Diaz2022,Bianchi2022}). 
More specifically, a bright arc is observed from the position of VLA4A extending towards the south-east. Other two arcs are 
extending from VLA4B towards the east and the west side of VLA4A. In addition, we detect three elongated structures \textbf{which} extend from the northern edge of the continuum associated with VLA4A and B towards the north. For simplicity, further in the text we refer to the central arc-like structure as the northern streamer.
In order to verify that these structures are not artifacts introduced by the cleaning process, we test different beam tapering.
All the elongated arc-like structures are still present when using a larger circular beam. We conclude that the dust emission is tracing accretion streamers extending towards the north and the south and feeding VLA4A and VLA4B.

%===Figura continuo con quadretti nel finger

%\begin{figure*}[h]
%\centering
%  \includegraphics[height=10cm]{Positions.eps}
%  \caption{Continuum emission as observed by ALMA at 1.2 mm (color scale and contours). First contours (gray) and steps (black) are 8$\sigma$, corresponding to 3.35 mJy beam$^{-1}$. The beam size is 0$\farcs$19 $\times$ 0$\farcs$13 (PA=+5$^{\circ}$). The binary system is composed of VLA4A (right) and VLA4B (left). Coordinates are $\alpha_{\rm J2000}$ = 03$^{\rm h}$ 29$^{\rm m}$ 3$\fs$75, $\delta_{\rm J2000}$ = +31$\degr$ 16$\arcmin$ 03$\farcs$81 and $\alpha_{\rm J2000}$ = 03$^{\rm h}$ 29$^{\rm m}$ 3$\fs$773, $\delta_{\rm J2000}$ = +31$\degr$ 16$\arcmin$ 03$\farcs$81 for VLA4A and VLA4B, respectively. The white squares indicate the offset positions where the spectra in Fig. have been extracted.}
%  \label{fig:Continuum}
%\end{figure*}

The optical source position measured using \textit{Gaia} is very close to VLA4B, even though there is a non-negligible discrepancy suggesting that the protostar is very embedded and the optical emission is due to light scattered by the circumbinary material\cite{Diaz2022}. The source driving the molecular jet, as well as a microjet traced by [Fe II], has also been identified with VLA4B \cite{Lefevre2017,Hodapp2014}.
When observed at mm wavelengths, VLA4B is brighter than VLA4A of a factor 3.8 at 7 mm and 3.7 at 9mm \cite{Anglada2000}, while the flux becomes comparable at longer wavelengths\cite{Tobin2016,Tyc2018}. The situation reverses at 1.2 mm where VLA4A is brighter by a factor of 1.5\cite{Bianchi2022}. The analysis of SEDs reveals the presence of free-free emission for both VLA4A and VLA4B, consistent with the presence of radio jets\cite{Diaz2022}.
The different spectral indices measured toward the sources, as well as the change in intensities, suggest that the dust at 1.2 mm is optically thick in VLA4A and very optically thick in VLA4B\cite{Tyc2018,Diaz2022}.
In this context, VLA4B is interpreted as the primary and more massive component of the binary system, while VLA4A would be the less evolved secondary component.
Taking into account the high uncertainties related to the free-free contamination, the disk masses have been estimated to be $\sim$0.2 M$_{\odot}$ for VLA4B and $\sim$0.08 M$_{\odot}$ for VLA4A\cite{Tyc2018}, confirming this picture.
Interestingly, SVS13-A has been also deeply investigated in the NIR and found to be in a phase of active accretion\cite{Fiorellino2021}. The quick rise of magnitude in the \textit{K}-band, followed by a slow decline, is reminiscent of FU Ori sources while the small bursts amplitude and the line rich spectra resemble EX Ori sources\cite{Hodapp2014}.

%==Figura continuum Bianchi2022

\subsection{The prototype Class I hot corinos}
The SVS13-A has been found to host a hot corino chemistry. More specifically, young Class 0 protostars possess an inner region, typically $\leq$ 100 au, where the gas temperature is high enough ($\geq$ 100 K) to desorbe the icy mantles of dust grains and release into the gas phase complex molecules which can in turn react in the gas-phase and form more complex molecular species. The hot corino phenomenon was first observed in Class 0 protostars (e.g. IRAS16293–2422 \cite{Cazaux2003}) and only recently it was also discovered in more evolved Class I objects. In this respect, SVS13-A provided the first evidence that Class I objects can be as chemically rich as young Class 0 sources \cite{Bianchi2019}. 
More specifically, water and iCOMs emission have been detected from a compact ($<$ 100 au) and hot ($>$ 100 K) region\cite{Codella2016,DeSimone2017,Bianchi2019} towards the source. 
Since then the chemistry of the system was deeply investigated thanks to several observational Large Programs including ASAI (Astrochemical Surveys At IRAM 30m; \cite{Lefloch2018}), CALYPSO (Continuum and Lines in Young ProtoStellar Objects; \cite{Belloche2020}) with the Plateau de Bure interferometer, SOLIS (Seed Of Life In Space; \cite{Ceccarelli2017}) with IRAM-NOEMA and PEACHES (Perseus ALMA Chemistry Survey; \cite{Yang2021}).
These projects identified more than 100 emission lines emitted by iCOMs including CH$_{\rm 3}$OH and its isotopologues, CH$_{\rm 3}$CHO, HCOOCH$_{\rm 3}$, CH$_{\rm 3}$OCH$_{\rm 3}$, HCOCH$_{\rm 2}$OH, NH$_{\rm 2}$CHO, CH$_{\rm 3}$CH$_{\rm 2}$OH. 
The broad spectral coverage of these surveys allowed a detailed multi line analysis, which results in an accurate determination of the gas properties. In particular, rotational temperatures and column density of each molecule have been determined and compared to those measured in other objects. The relative iCOMs abundances are in good agreement both with measurements in Class 0 protostars and to those in the 67P/Churyumov–Gerasimenko comet, suggesting to some extent inheritance of complexity from the early stages \cite{Bianchi2019}.
ALMA high-angular resolution observations ($\sim$ 50 au) shed new light on the distribution of iCOMs across the binary system. In particular, a striking chemical segregation was observed. Some molecular species, such as methanol and dimethyl ether, are detected in both VLA4A and VLA4B while others, e.g. formamide and ethylene glycol are detected only in VLA4A \cite{Bianchi2022,Diaz2022}. Interestingly, the line emission appears to be elongated towards the north (up to $\sim$150 au), where one of the accretion streamers in the dust is present (see Fig. \ref{fig:Continuum}). The origin of such molecular line distribution needs to be clarified using different molecular tracers. To this purpose we report here the analysis of sulfur dioxide SO$_2$ and deuterated water HDO in the SVS13-A system.

\section{Observations}\label{sec:observations}

\begin{figure*}[h]
\centering
  \includegraphics[width=\textwidth]{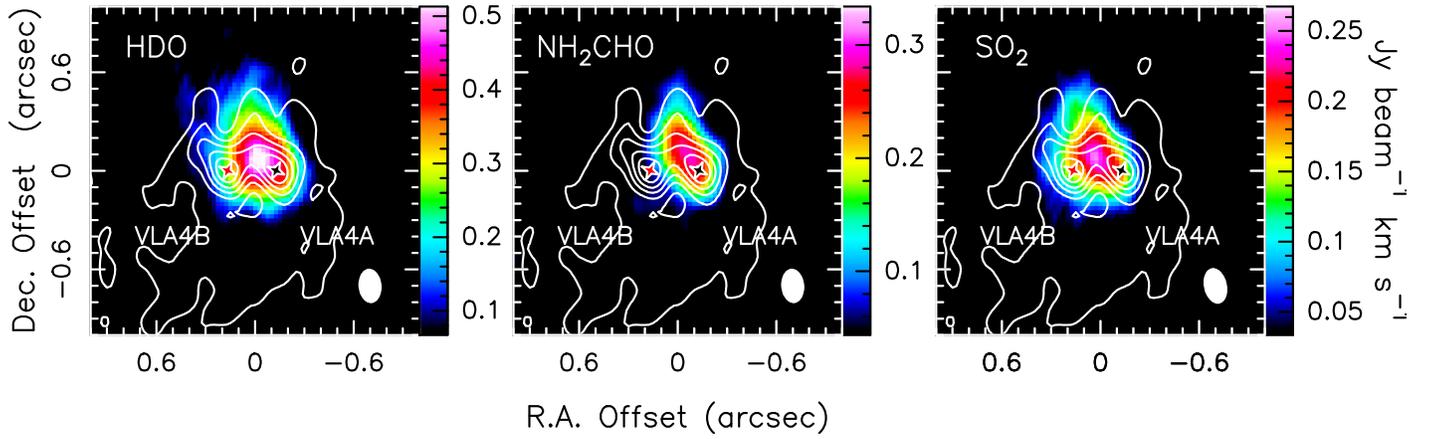}
  \caption{Velocity integrated emission (moment 0) of the HDO 2(1,1)--2(1,2) (left panel), NH$_{\rm 2}$CHO 12(1,12)--11(1,11) (middle panel) and SO$_2$ 5(2,4)--4(1,3) (right panel) lines, in colour scale, superposed to the dust emission in white contours. The synthesised beams are reported in white in the lower right corner of each panel. The black and red stars indicate the VLA4A and VLA4B positions, respectively.}
  \label{fig:mom0}
\end{figure*}

The observations presented here have been acquired using the Atacama Large Millimeter/submillimeter Array (ALMA) on 2019 September 8 during the project (2018.1.01461.S, see also \cite{Bianchi2022}). The telescope was in the C43-6 configuration, with baselines between 43 m and 5.9 km. 
The center of the observed field is 
$\alpha_{\rm J2000}$ = 03$^{\rm h}$ 29$^{\rm m}$ 3$\fs$8,
$\delta_{\rm J2000}$ = +31$\degr$ 16$\arcmin$ 03$\farcs$8.
The bandpass and flux calibrator was the quasar J0510+1800, while the phase calibrator was J0336+3218. 
Data calibration was performed using the standard ALMA calibration pipeline in \textsc{CASA} \citep{McMullin2007}.
Successively, we used the \textsc{IRAM-GILDAS}\footnote{\url{http://www.iram.fr/IRAMFR/GILDAS}} package to determine line-free continuum channels, perform phase self-calibration and apply the solution both to the continuum and the spectral cubes.
The final calibration uncertainty on the absolute flux is 20\%.

\section{Results}\label{sec:results}

\begin{figure*}[h]
\centering
  \includegraphics[height=11cm]{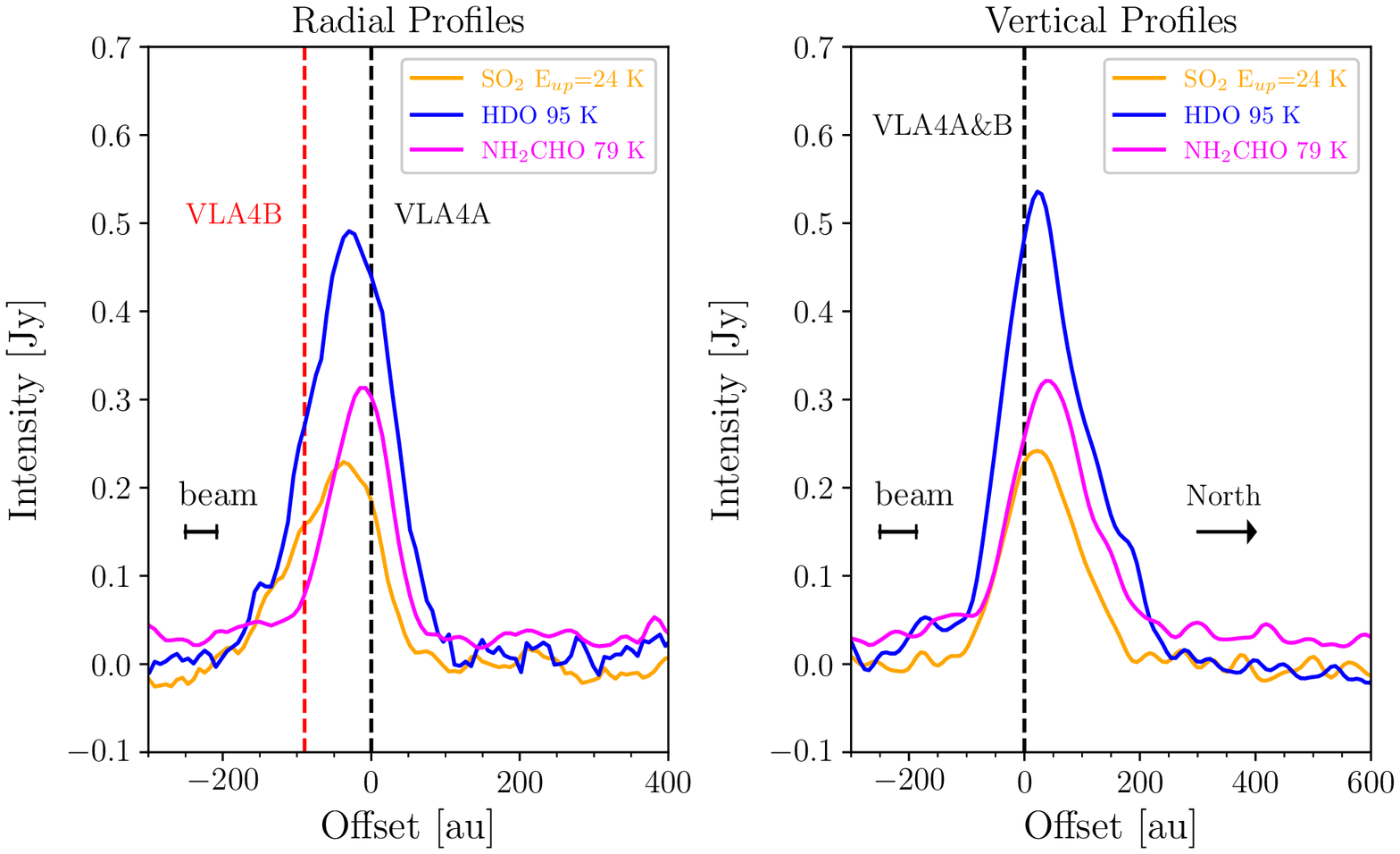}
  \caption{\textit{Left panel:} Intensity profile of the HDO line at 241.56155 GHz (blue) and the SO$_2$ line at 241.615797 GHz (orange) extracted along the line connecting VLA4A with VLA4B. The two vertical
dashed lines show the position of VLA4A (black) and VLA4B (red). \textit{Right panel:} Intensity profile of the same lines extracted along the vertical line crossing the northern streamer. The vertical dashed lines show the position of VLA4A and VLA4B.}
  \label{fig:profiles}
\end{figure*}

\begin{figure*}[h]
%\centering
  \includegraphics[height=5cm,right]{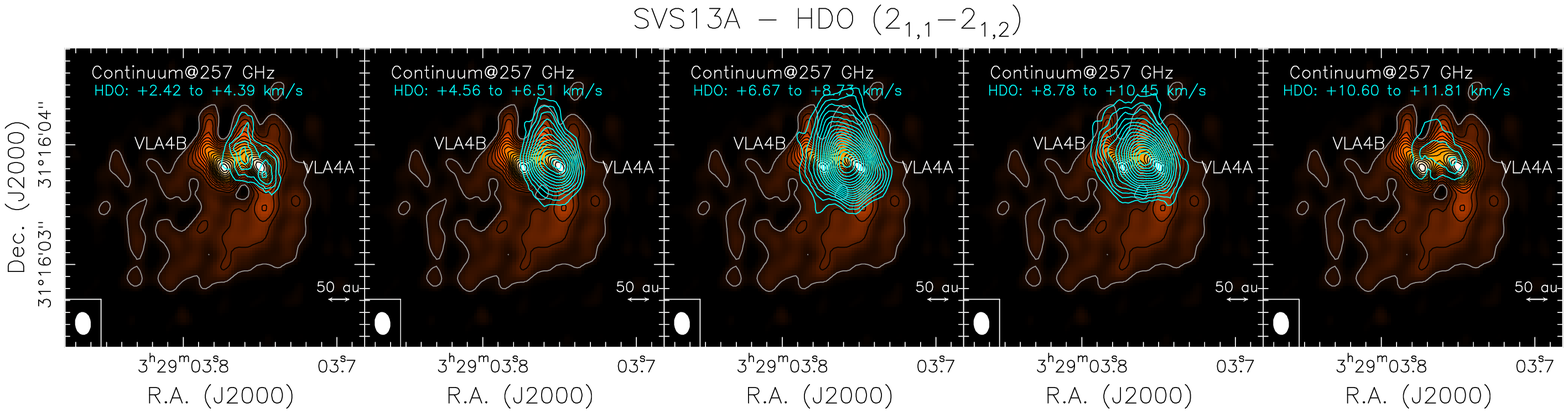}
  \includegraphics[height=5cm,right]{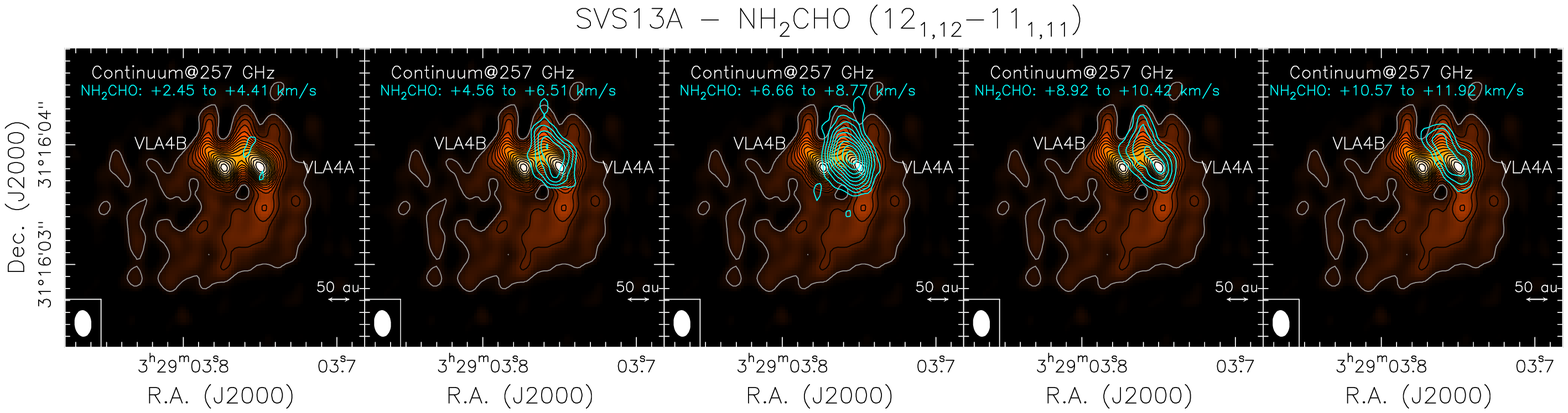}
  \includegraphics[height=5cm,right]{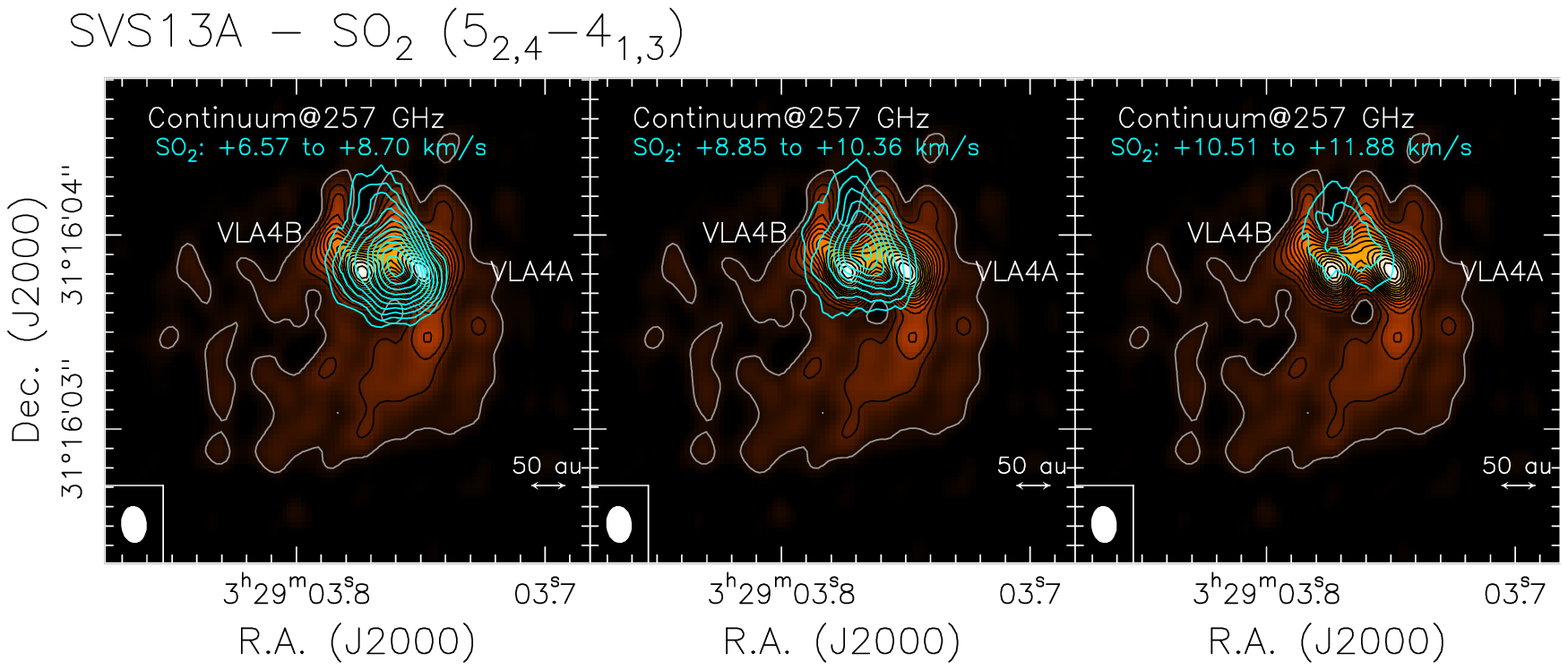}
  \caption{{\it Upper panels:} Channel map of the HDO(2$_{\rm1,1}-2{\rm ,2}$) emission (cyan contours) overlapped on the 1.2 continuum map of SVS13-A VLA4A and VLA4B (see Fig. \ref{fig:Continuum}). Five velocity intervals have been considered (reported in the Top-Left corners). First contours and steps correspond to 5$\sigma$ (20 mJy beam$^{-1}$ km s$^{-1}$ for all the channels but the most blue- and red-shifted ones, 16 mJy beam$^{-1}$ km s$^{-1}$), and 2$\sigma$, respectively. 
  The systemic velocities are: +7.7 km s$^{-1}$ (VLA4A), and
  +8.5 km s$^{-1}$ (VLA4B). The synthesised beam (Bottom-Left) is 207 $\times$ 136 mas (PA = +5$\degr$). 
  {\it Middle panels:} same as Upper for NH$_2$CHO (12$_{\rm 1,12}-11_{\rm 1,11}$). First contours and steps as for HDO.
  {\it Lower panels:} same as Upper for SO$_2$ (5$_{\rm 2,4}-4_{\rm 1,3}$). 
  First contours and steps as for HDO.
  In this case, only three velocity intervals can be drawn because of the contamination due to the CH$_{\rm 2}$DOH(4$_{\rm 3,1}-5_{\rm 1,5}$)e1 line at 241.61928840 GHz, i.e.
  at --4.3 km s$^{-1}$ from the SO$_2$ rest line (see Table \ref{table:lines}).}
  \label{fig:channels}
\end{figure*}

We detect and analyse the HDO 2(1,1)--2(1,2) transition at 241.561550 GHz with an upper level energy (E$_{\rm up}$) of 95 K. In the same spectral window we also detect the SO$_{\rm 2}$ 5(2,4)--4(1,3) transition at 241.615797 GHz with E$_{\rm up}$ of 24 K. The line spectroscopic parameters are listed in Table \ref{table:lines}.
The rms noise is typically 4.8 mJy beam$^{-1}$ km s$^{-1}$ using spectral channels of 122 kHz (0.15 km s$^{-1}$).
%We also detected the HDO 7(3,4)--6(4,3) transition at 241.973570 GHz with E$_{\rm up}$= 837 K with a rms noise of 4.9 mJy beam$^{-1}$ km s$^{-1}$ in the same spectral channels.
The synthesized beam is 0$\farcs$21 $\times$ 0$\farcs$14 for both the line cubes.
The analysis is supported by the NH$_2$CHO 12(1,12)--11(1,11) transition at 243.521044 GHz, previously published in \cite{Bianchi2022}. 
 %In this case, the rms noise is 6.2 mJy beam$^{-1}$ km s$^{-1}$ using spectral channels of 122 kHz (0.15 km s$^{-1}$).

\begin{table*}[tbh]
  \caption{List of transitions and line properties (in T$_{\rm MB}$ scale) of the HDO, SO$_2$ and NH$_{\rm 2}$CHO emission. The columns report the transition and their frequency (GHz), the upper level energy E$_{\rm up}$ (K), the line strength $S\mu^2$ (D$^2$), the velocity integrated line intensity I$_{\rm int}$ (mK km s$^{-1}$) and the intensity ratio.}
  \label{table:lines}
  \begin{tabular*}{1.0\textwidth}{@{\extracolsep{\fill}}llcccccc}
    \hline
Species & Transition & $\nu$$^{\rm a}$ & E$_{\rm up}$$^{\rm a}$ & $S\mu^2$$^{\rm a}$ & \multicolumn{2}{c}{I$_{\rm int}$$^{\rm b}$}& Intensity ratio\\
%	\hline
 & & (GHz) & (K) & (D$^2$) & \multicolumn{2}{c}{(K km s$^{-1}$)}&\\
\hline
&&&&& VLA4A & VLA4B& VLA4A/VLA4B\\
\hline
%1(1,0)-1(1,1) & 80.578295 & 47 & 0.65 & I & I & ratio unresolved\\
HDO & 2(1,1)--2(1,2) & 241.561550 & 95 & 0.4 & 349 (70)  & 177 (36) & 2.0 (0.6) \\
%HDO & 7(3,4)--6(4,3) & 241.973570 & 837 & 1.39 & 179 (36) & 94 (19) & 1.9 (0.5) \\
%\hline
SO$_2$$^{\rm c}$ & 5(2,4)--4(1,3)& 241.615797 & 24 & 5.7 & $\geq$126 & $\geq$ 175 & - \\
%& 126 (25) & 175 (35) & - \\

NH$_{\rm 2}$CHO$^{\rm d}$ & 12(1,12)--11(1,11) & 243.521044  & 79 & 156 & 267 (59) & 40 (11) & 7 (2)\\

%NH$_{\rm 2}$CHO$^{\rm d}$& 12$_{\rm 4,9}$--11$_{\rm 4,8}$ & 255.058533  & 127 & 139 & 160 (37) & 42 (12) & 4 (1) \\
    \hline
%$^{13}$CH$_3$OH & 4(3,2)--4(2,3)A & 255.203728 & 73 & 138 (31) & 102 (23) & 1.4 (0.4)\\
%\hline 
%CH$_{\rm 3}$OH 25$_{\rm 3,22}$--25$_{\rm 2,23}$ A& 241.58876 & 804 & 251 (53) & 112 (26) & 2.2 (0.7) \\
%\hline
  \end{tabular*}
  $^{\rm a}$ Frequencies and spectroscopic parameters have been provided by
\cite{Messer1984} for HDO, and retrieved from the Jet Propulsion Laboratory molecular database (https://spec.jpl.nasa.gov/) \cite{Pickett1998}. For SO$_2$ spectroscopic parameters are provided by the Cologne Database for Molecular Spectroscopy (http://www.astro.unikoeln.de/cdms/) \cite{Muller2005}. For NH$_2$CHO spectroscopic parameters are provided by \cite{Kukolich1971} and retrieved from the CDMS database.
$^{\rm b}$ Errors on the integrated intensity include 20$\%$ of calibration. 
$^{\rm c}$ Contaminated by the CH$_{\rm 2}$DOH 4$_{\rm 3,1}$--5$_{\rm 1,5}$ e1 line at 241.61928840 GHz, at --4.3 km s$^{-1}$ from the SO$_2$ rest line.
$^{\rm d}$ From \cite{Bianchi2022}. 
\end{table*}

Figure \ref{fig:mom0} shows the spatial distribution of the HDO (left panel), NH$_2$CHO (middle panel) and SO$_2$ (right panel) lines with respect to the dust continuum emission. In particular, the line emission is integrated over the whole velocity interval and reported in color scale superposed to the continuum emission in white contours. As already noticed in \cite{Bianchi2022}, the NH$_2$CHO emission is more compact and centered in VLA4A while HDO and SO$_2$ have a broader emission, extended also towards VLA4B. 
This finding confirms the chemical segregation reported in the binary system using iCOMs \cite{Bianchi2022,Diaz2022}. Moreover, HDO and SO$_2$ emission show an elongation toward the north, as already observed for iCOMs. 

% Fig. 3 Radial and Vertical profiles

In order to further investigate the lines, radial and vertical distribution, we plot in Fig. \ref{fig:profiles} the intensity [Jy] profiles of SO$_2$, HDO and NH$_2$CHO extracted from the moment 0 maps (see Fig. \ref{fig:mom0}) along a horizontal axis connecting the two protostars (left panel) and a vertical axis between the protostars crossing the northern streamer (right panel).
The x-axis reports the offset in astronomical units with respect to the position of VLA4A. The vertical dashed lines indicate the positions of the two protostars.
The radial profiles show that HDO and SO$_2$ are extended in a region covering both protostars with a peak of emission in between. Despite the difference in the peak intensity, there are not substantial differences between the spatial distribution of HDO and SO$_2$. Conversely, NH$_2$CHO has the peak of the emission closer to VLA4A\cite{Bianchi2022}.
In addition, the vertical profiles show that all the molecules are extended toward the north direction at a larger distance than the dust streamer, located at about $\sim$ 0$\farcs$4 (corresponding to 120 au at the source distance). The radial and vertical profiles do not show a trend with the upper level energy of the different lines, confirming that the molecular distribution do not strongly depend on the excitation conditions\cite{Bianchi2022}.

We report in Table \ref{table:lines} the integrated line intensities extracted from the continuum peak positions of VLA4A and VLA4B, respectively, and their intensity ratios. The intensity ratio for HDO is 2.0 (0.6), suggesting that water is distributed in both the protostars, even if it is more abundant in VLA4A. 
Similar intensity ratios were derived for iCOMs such as CH$_3$OH, and CH$_3$CHO\cite{Bianchi2022}.
On the other hand, the intensity ratio measured for the NH$_2$CHO line is 7 (2), confirming the presence of formamide predominantly towards VLA4A. 
We also reported the intensity of SO$_{\rm 2}$ in VLA4A and VLA4B. The ratio is not reported in this case since the blue-shifted part of the line is contaminated and the line intensity extracted in VLA4A is a lower limit. The HDO line intensity in the brighter source, VLA4A, is consistent with the rotational temperature of 150-260 K and a column density of 4 $\times$ 10$^{17}$ cm$^{-2}$, derived using IRAM-30m data collecting the emission from both protostars\cite{Codella2016}.  
The NH$_2$CHO rotational temperatures and column densities derived in \cite{Bianchi2022}, are T$_{\rm rot}$=140 K and N$_{tot}$=5.9(0.9) $\times$ 10$^{15}$ cm$^{-2}$ in VLA4A and T$_{\rm rot}$=170 K and N$_{tot}$=1.4(0.3) $\times$ 10$^{15}$ cm$^{-2}$ in VLA4B.

Figure \ref{fig:channels} shows the spatial distribution of HDO (upper panels), NH$_2$CHO (middle panels) and SO$_2$ (lower panels) with respect to the dust continuum emission. More specifically, line emission is integrated in five different velocity channels and showed in cyan contours superposed to the continuum emission in color scale.
The NH$_2$CHO emission in all the velocity intervals traces only VLA4A and it is elongated toward the northern streamer. 
For HDO the blue-shifted emission in the velocity intervals 2.4-4.4 km s$^{-1}$ and 4.6-6.5 km s$^{-1}$ shows the same distribution.
In the velocity intervals 6.7-8.7 km s$^{-1}$, 8.8-10.5 km s$^{-1}$ and 10.6-11.8 km s$^{-1}$ the line emission is
associated with both protostars, peaking in between.
The central panels correspond to the systemic velocities of VLA4A and VLA4B, which are +7.7 km s$^{-1}$ and +8.5 km s$^{-1}$, respectively, and have been determined from methanol emission\cite{Bianchi2022}.
SO$_2$ emission is also associated to both protostars in the velocity intervals 6.6-8.7 km s$^{-1}$, 8.9-10.4 km s$^{-1}$, and 10.5-11.9 km s$^{-1}$. Even if the blue-shifted part of the line (velocities lower than +6.5 km s$^{-1}$) is contaminated by the CH$_{\rm 2}$DOH(4$_{\rm 3,1}$--5$_{\rm 1,5}$)e1 line at 241.61928840 GHz (see Table \ref{table:lines}), the SO$_2$ emission in the other available channels shows a spatial distribution similar to that of HDO.
The overall line distribution confirms what previously reported by \cite{Bianchi2022} but the HDO emission highlights an additional structure (i) with a peak spatially shifted with respect to the protostars position, and (ii) at velocities blue-shifted by $\geq$ 3 km s$^{-1}$. This component is perfectly aligned with the northern streamer traced by the dust, thus revealing molecular gas associated with the streamer itself.
Interestingly, in the same blue-shifted velocity range, NH$_2$CHO shows hints of this component even if the S/N is lower than that of HDO (see Fig. \ref{fig:profiles}).

%In Fig. \ref{fig:spectra} we report the HDO spectra extracted at the continuum position of VLA4A (top panel), VLA4B (middle panel) and integrated in a region of 0$\farcs$1 $\times$ 0$\farcs$1 around the northern streamer, namely at offset larger than 300 mas from the field center. The dashed lines in the top and middle panels indicate the systemic velocities of VLA4A (+7.7 km s$^{-1}$) and VLA4B (+8.5 km s$^{-1}$), respectively.
%In the bottom panel, a peak of emission is detected at the blue-shifted velocity of +2 km s$^{-1}$. Spectra are smoothed to a velocity resolution of 0.5 km s$^{-1}$ in order to increase the signal-to-noise.

%================================================================

%\begin{figure*}[h]
%\centering
%  \includegraphics[height=12cm]{HDO-spw1.eps}
%  \caption{ALMA observations of the HDO}
%  \label{fig:HDO-SO-maps}
%\end{figure*}

%%%%%%%%%%%%%%%%%%%%%%%%%%%%%%%%%
\section{Discussion}\label{sec:discussion}
\subsection{Binding energies and sublimation temperatures}\label{subsec:chemistry}
As we will discuss in more detail in the next subsection, the presence in the gas-phase of deuterated water, SO$_2$ and formamide provides new insights into their origin.

First, deuterated water is prevalently formed in the cold prestellar phase on the grain surfaces, in the so-called mantle (e.g. \cite{Ceccarelli2014}). 
Therefore, HDO remains frozen unless it is released into the gas-phase by some process. 
In general, two processes can possibly inject HDO into the gas-phase (e.g. see the discussion in \cite{Ceccarelli2022}): the thermal sublimation from warm dust grain or a (non-thermal) shock.
Other possible mechanisms, such as chemical desorption or photo-desorption, would not be efficient here.
About formamide, there is some discussion on its formation route.
In principle, it can be synthesised on the grain surfaces (e.g. \cite{Garrod2008}) or in the gas-phase \cite{Skouteris2017} from reactants (H$_2$CO and NH$_2$) that are formed on the grain surfaces, or both.
If formed on the grain surfaces, formamide would need to be released into the gas-phase by either thermal sublimation or a (non-thermal) shock, as HDO.
The same consideration applies if formamide is formed in the gas-phase, because the two reactants H$_2$CO and NH$_2$ would need to be in the gaseous form.
Finally, SO$_2$ is formed in the gas-phase, but Sulphur is observed to be depleted, by about a factor 20, in the cold molecular clouds \cite{Fuente2023}, so that it is often taken as a tracer of warm and shocked regions.
In summary, the presence of HDO, SO$_2$ and NH$_2$CHO, implies that either they (or their precursors) are thermally sublimated or they are injected from the frozen mantles into the gas-phase by a shock.

In order to distinguish which of the two hypothesis is correct, one has to evaluate the thermal sublimation temperature of deuterated water, SO$_2$ and formamide.
To this end, we used the half-life time $t_{sub}$ of the frozen species:
\begin{equation}\label{eq:sub_time}
    t_{sub}=ln(2)/k_{\rm des}
\end{equation}
The rate of desorption $k_{\rm des}$ is given by the usual equation:
\begin{equation}\label{eq:desorption}
    k_{\rm des} = \nu_{des} \exp({-BE/T}) 
\end{equation}
where BE is the binding energy, given in units of temperature, and $\nu_{des}$ is the pre-exponential factor, which depends on the adsorbate and surface (e.g. \cite{Minissale2022,Ferrero2022}).

Table \ref{tab:BE} list the BE of HDO, SO$_2$ and NH$_2$CHO over an amorphous water surface, as computed by \cite{Ferrero2020} and \cite{Perrero2022} on a ice model of 60 water molecules.
Note that the HDO BE was assumed to be equal to that of H$_2$O, an assumption demonstrated to be valid by new computations of Tinacci et al. (submitted) on a large (200 molecules) water ice grain.
The table also reports the pre-exponential factor $\nu$ of H$_2$O and NH$_2$CHO, computed by \cite{Minissale2022} following the \cite{Tait2005} formalism (Table 5 of \cite{Minissale2022}).
The pre-exponential factor of SO$_2$ was estimated by us using the formalism by \cite{Tait2005}.
\begin{table}[]
    \centering
    \begin{tabular}{ccc}
        \hline
        Species & BE [K] & $\nu$ [s$^{-1}$] \\
        \hline
         HDO & 3605--6111 & $5\times 10^{15}$\\
         SO$_2$ & 2105 -- 5827 & $2\times 10^{16}$\\
         NH$_2$CHO & 5793--10960 & $4\times 10^{18}$\\
        \hline
    \end{tabular}
    \caption{Binding energies (BE) and pre-exponential factor ($\nu$) used for deriving the sublimation temperatures shown in Fig. \ref{fig:Tsub}.
    The BE are from \cite{Ferrero2020} and \cite{Perrero2022}.
    The pre-exponential factors are from \cite{Minissale2022} (see text).
    Note that the second column reports minimum and maximum values of BE computed on a amorphous water surface model of 60 molecules.
    Third column reports the pre-exponential factor computed using the \cite{Tait2005} method, also described and discussed in \cite{Minissale2022} and \cite{Ferrero2022}.}
    \label{tab:BE}
\end{table}

\begin{figure}
    \centering
    \includegraphics[width=8.5cm]{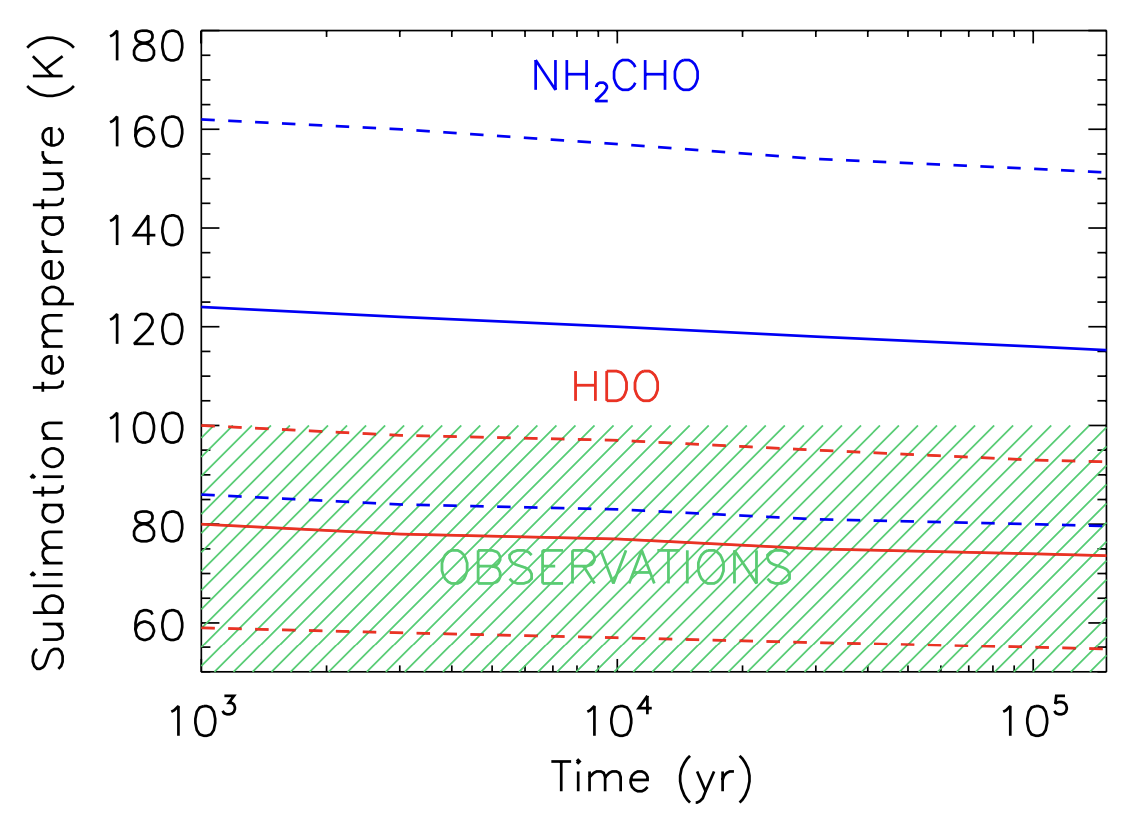}
    \caption{Sublimation temperature of deuterated water (HDO; red lines) and formamide (NH$_2$CHO; blue lines). 
    The three curves for each species correspond to the average (solid line), minimum and maximum (dashed lines) BE computed by \cite{Ferrero2020} and reported in Table \ref{tab:BE} (see text).
    Note that for the HDO we used the same BE of H$_2$O, as demonstrated to be valid by Tinacci et al. (submitted).
    The green area shows the grain temperature derived by \cite{Bianchi2022b} assuming a bolometric luminosity of 55 L$_o$ and a distance from the center of 120 au (see text).}
    \label{fig:Tsub}
\end{figure}

Figure \ref{fig:Tsub} shows the derived sublimation temperature of deuterated water and formamide, for a time interval between $10^3$ and $10^5$ yr.
We do not show the case of SO$_2$ for clarity reasons, since its sublimation temperature would lie below that of HDO.
The figure shows three sublimation temperature curves for each species, corresponding to the average, minimum and maximum BE computed by \cite{Ferrero2020}.
Assuming that, in reality, BE has a distribution of values approximately described by a gaussian (e.g. \cite{Tinacci2022}), the curve corresponding to the average would represent where most of the relevant molecules would thermally sublimate.
While the HDO being in the gaseous form is, in principle (see below) compatible with thermal sublimation, formamide is not.
The implication of this fact will be discussed in the following subsection.

%%%%%%%
\subsection{Accretion streamers feeding the disks}

The bolometric luminosity (L$_{bol}$) of VLA4A and VLA4B is not measured singularly due to angular resolution limitations at infrared wavelengths.
The total bolometric luminosity of VLA4A+VLA4B is estimated to be in the range 45-55 L$_{\odot}$\cite{Tobin2016,Podio2021} at the source distance of 299 pc. The dust streamer traced by the continuum emission extends at a distance larger than 0$\farcs$4 (120 au). 
If we consider a spherical envelope-like distribution, and conservatively a bolometric luminosity of 55 L$_{\odot}$, the temperature at this distance from the protostars is lower than 100 K (see e.g. \cite{Ceccarelli2000}).
The presence of formamide at this position, considering the information on the binding energies and the corresponding sublimation temperatures (see \ref{subsec:chemistry} and Fig. \ref{fig:Tsub}), suggests that formamide is not thermally desorbed.
In the case of HDO and SO$_2$ the binding energies are lower so their presence cannot exclude thermal desorption.
If the continuum emission around the protostars originates from an inclined disk (instead of a spherical envelope), as suggested by the observed geometry \citep{Diaz2022}, the snowlines of NH$_2$CHO, HDO, and SO$_2$, would be even closer to the source. Therefore the detection of molecules at a distance of $\geq$ 120 au, would exclude the thermal evaporation scenario. 

However, episodic accretion is a common process in young stellar objects.
More specifically, observations performed using the Very Large Telescope, indicate that SVS13-A is in a phase of active accretion, deriving a mass accretion rate of a few 10$^{-6}$ M$_{\odot}$ yr$^{-1}$\cite{Fiorellino2021} (see also \cite{Hsieh2019}). 
The presence of an accretion burst is in principle expected to move the snowlines outwards and inject into the gas phase the molecules stored in the dust mantles, as in the case of the Fu Ori V833 disk\cite{Lee2019}.
However, at near-IR wavelengths VLA4A is not detected\cite{Hodapp2014}, suggesting a negligible contribution to the bolometric luminosity which will be instead dominated by VLA4B\cite{Fiorellino2021}.
This is consistent with VLA4B being the primary and more massive
component of the binary system, while VLA4A would be the less massive secondary component.
The process of differential accretion in a binary system has been studied using numerical simulations and the secondary is expected to accrete most of the gas, except in the case of very eccentric orbits (e.g. \cite{Bate1997, Ceppi2022}).
In this scenario VLA4A would be the component actively accreting, and the associated accretion burst may be at the origin of the detected molecular emission associated with VLA4A. In order to definitely exclude the possibility that thermal sublimation is responsible for the chemical enrichment at distances larger than 120 au, it is crucial to have an estimate of the individual VLA4A and VLA4B luminosities. 

Alternatively, the emission may be produced by an accretion shock, sputtering the grains, occurring when the accretion streamer impacts the disk of VLA4A. This would also justify the fact that the molecular emission shows emission at both blue-shifted and red-shifted velocities in the streamer (see Fig. \ref{fig:channels}), which may be due to the velocity dispersion caused by the shock. Indeed, a similar scenario has been proposed to explain SO, and SO$_2$ emission where molecular streamers impact onto the disks in Class I/II sources observed by the ALMA-DOT project \cite{Garufi2022}. Also in these sources we observe both blue-shifted and red-shifted SO, and SO$_2$ emission at the impact region.

Recently, the possible presence of \textbf{accretion shocks or} self-gravitating substructures (see e.g. \cite{Kratter2016}) has been invoked to explain local temperature enhancements in the circumbinary material of the Class 0 source IRAS16293-2422 \cite{Maureira2022}. However, in the case of SVS13-A the circumstellar disks around VLA4A and VLA4B, as well as the circumbinary material are expected to be stable against local gravitational instabilities, as found by \cite{Diaz2022}.

In summary, a major result of the presented observations is the detection of molecular gas in the accretion streamer feeding the system SVS13-A, on scales of $\sim$ 100-200 au. 
 Interestingly, an accretion streamer connecting the source to the outer envelope, up to $\sim$ 700 au has been recently imaged in DCN using the IRAM NOEMA interferometer\cite{Prodige2023}.
 The detected DCN streamer is associated with the southern dusty streamer, even though some emission is present also in the northern region (see their Fig. C.1). However, the limited angular resolution of the NOEMA observations (synthesized beam of 1$\farcs$2 $\times$ 0$\farcs$7) prevents a proper comparison with the ALMA continuum emission.
Finally, the present observations confirm that deuterated water and iCOMs, including formamide, are as effective as classical tracers such as SO and SO$_2$, in revealing and investigating accretion processes in young embedded Class 0/I disks.

%\begin{comment}
%\subsection{Missing water in Class I?}
%Class I disks have been found to be poor in water (Harsono et al. 2020). However SVS13-A is a counter example in which water is very abundant, at least HDO.
%They say that probably they do not see water because it comes from the disk region and it is obscured by the dust (optical depth.) Our observations confirm that water emission comes from the disk in Class I.
%Their targets are TMC1A, L1527 IRS in Taurus and Elias 29, GSS 30 IRS 1, and GSS 30 IRS 3 in $/rho$-Ophiuchi. 
%they suggest that in this objects water if present is in the inner 10 au, while the envelope is dry.
%Our observations indicate that it is more extended.
%Other differences? iCOMs detected in their sources? Possible explaination?
%\end{comment}

\section{Conclusions}\label{sec:conclusions}

We report new ALMA observations of HDO and SO$_2$ at a $\sim$ 50 au angular resolution of the protobinary system SVS13-A. 
The line spatial distribution, as well as the radial and vertical profiles
indicates that HDO and SO$_2$ are associated with both the two components of the binary system, VLA4A and VLA4B. 
We compare the spatial distribution with that of NH$_2$CHO, which instead mainly trace the VLA4A component, confirming a chemical differentiation 
around the two protostars.
In addition, the channel maps of HDO emission reveal a further component at blue-shifted velocities (+2.42--4.39 km s$^{-1}$) spatially coincident with the dust accretion streamer. The streamer traced by HDO is located at a distance $\geq$ 120 au from the VLA4A and VLA4B protostars. Formamide is also detected at the same spatial position.
In order to investigate the origin of the molecular emission in the streamer, we evaluate the thermal sublimation temperatures of deuterated water, SO$_2$ and formamide, using updated BE distributions.
We find that a thermal sublimation origin of the molecular emission is excluded, unless the VLA4A source is currently experiencing an accretion burst. Alternatively, the emission morphology is consistent with an accretion shock, produced by an accretion streamer impacting the VLA4A disk, as previously found in more evolved protoplanetary disks.
Finally, water has a key role in the planet formation process and it is relevant to investigate its abundance across the different evolutionary stages.
The spatial distribution of HDO in SVS13-A demonstrates that in Class I disks water is not necessarily confined in the inner 10 au, as suggested by previous ALMA and NOEMA H$_2^{18}$O observations in 5 Class I disks\cite{Harsono2020}.  
In this respect, accretion processes are a powerful tool to investigate the amount of water present in the young disks that will be eventually inherited from the forming planetary systems.

%You can also put lists into the text. You can have bulleted or numbered lists of almost any kind. 
%The \texttt{mhchem} package can also be used so that formulae are easy to input: \texttt{\textbackslash ce\{H2SO4\}} gives \ce{H2SO4}. 

%The conclusions section should come at the end of article. For the reference section, the style file \texttt{rsc.bst} can be used to generate the correct reference style.\footnote[4]{Footnotes should appear here. These might include comments relevant to but not central to the matter under discussion, limited experimental and spectral data, and crystallographic data.}

%\section*{Author Contributions}
%We strongly encourage authors to include author contributions and recommend using \href{https://casrai.org/credit/}{CRediT} for standardised contribution descriptions. Please refer to our general \href{https://www.rsc.org/journals-books-databases/journal-authors-reviewers/author-responsibilities/}{author guidelines} for more information about authorship.

\section*{Acknowledgements}
This project has received funding from: 
1) the Deutsche Forschungsgemeinschaft (DFG, German Research Foundation) under Germany´s Excellence Strategy – EXC 2094 – 390783311;
2) the European Research Council (ERC) under the European Union's Horizon 2020 research and innovation program, for the Project “The Dawn of Organic Chemistry” (DOC), grant agreement No 741002; 
3) the PRIN-INAF 2016 The Cradle of Life - GENESIS-SKA (General Conditions in Early Planetary Systems for the rise of life with SKA); 
4) the European Union’s Horizon 2020 research and innovation programs under projects “Astro-Chemistry Origins” (ACO), Grant No 811312.
This paper makes use of the following ALMA data: ADS/JAO.ALMA\#2018.1.01461.S. ALMA is a partnership of ESO (representing its member states), NSF (USA) and NINS (Japan), together with NRC (Canada), MOST and ASIAA (Taiwan), and KASI (Republic of Korea), in cooperation with the Republic of Chile. The Joint ALMA Observatory is operated by ESO, AUI/NRAO and NAOJ.
Most of the computations presented in this paper were performed using the GRICAD infrastructure (https://gricad.univ-grenoble-alpes.fr), which is partly supported by the Equip@Meso project (reference ANR-10-EQPX-29-01) of the programme Investissements d'Avenir supervised by the Agence Nationale pour la Recherche.\\
\textit{Software:} astropy \citep{Astropy1,Astropy2}, matplotlib \citep{Matplotlib}.

\section*{Conflicts of interest}
%In accordance with our policy on \href{https://www.rsc.org/journals-books-databases/journal-authors-reviewers/author-responsibilities/#code-of-conduct}{Conflicts of interest} please ensure that a conflicts of interest statement is included in your manuscript here.  Please note that this statement is required for all submitted manuscripts.  If no conflicts exist, please state that 
There are no conflicts to declare.
%%%END OF MAIN TEXT%%%

%  For footnotes in the main text of the article please number the footnotes to avoid duplicate symbols. e.g.  \footnote[num]{your text} the corresponding author \ast counts as footnote 1, ESI as footnote 2, e.g. if there is no ESI, please start at [num]=[2], if ESI is cited in the title please start at [num]=[3] etc. Please also cite the ESI within the main body of the text using \dag.

% The \balance command can be used to balance the columns on the final page if desired. It should be placed anywhere within the first column of the last page.

% \balance

% If notes are included in your references you can change the title from 'References' to 'Notes and references' using the following command:
% \renewcommand\refname{Notes and references}

%%%REFERENCES%%%
\scriptsize{
\bibliography{Mybib.bib} %You need to replace "rsc" on this line with the name of your .bib file
\bibliographystyle{rsc} } %the RSC's .bst file

\end{document}